\documentclass[10pt,conference,compsocconf]{./IEEEtran}
\usepackage[T1]{fontenc}
\usepackage{pslatex}
\usepackage{amsmath}
\usepackage{array}
\usepackage{booktabs}
\usepackage{cite}
\usepackage{epsfig}
\usepackage{flafter}
\usepackage{float}
\usepackage{tabularx}
\usepackage{color}
\usepackage{algorithm}
\usepackage{algpseudocode}

\begin{document}

\title{Simple {FPGA} routing graph compression}

\author{\IEEEauthorblockN{Andrew Kennings}
\IEEEauthorblockA{Electrical and Computer Engineering\\
University of Waterloo\\
Email: akennings@uwaterloo.ca}}

\maketitle
\begin{abstract}
Modern FPGAs continue to increase in capacity which requires
more memory to run the CAD flow.
The routing resource graph, which is needed by the detailed
router, is a memory hungry data structure
which describes all of the physical resources
and programmable connections within an FPGA.
We propose a compression scheme to reduce the memory 
requirements of the routing resource graph.
The scheme is simple to apply and requires only trivial
changes to the FPGA detailed routing algorithm.
The approach does not require any assumptions about the FPGA
routing architecture.  
Numerical results show excellent compression 
(\textcolor{black}{as much as $3.6$X overall memory reduction})
with only a slight increase ($\sim 20\%$ on average) on the router runtime
as a consequence of the routing graph compression.
\end{abstract}

\section{Introduction}

\begin{figure}[tb]
\begin{center}
\includegraphics[width=2.75in]{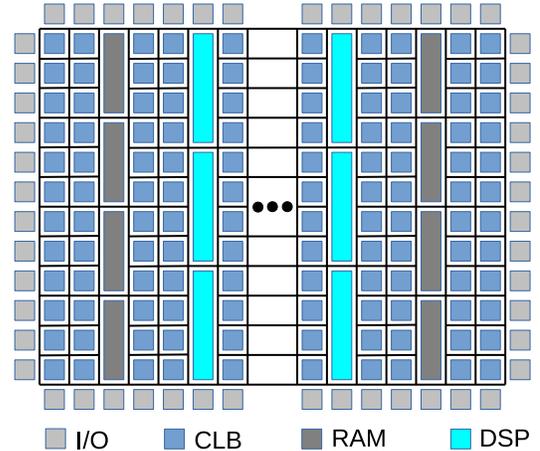} 
\end{center}
\vspace{-5mm}
\caption{Illustration of a modern two-dimensional 
island-styled heterogeneous FPGA consisting
of CLBs, DSPs, RAMs and I/Os.}
\label{fig:fpga}
\end{figure}

Modern Field Programmable Gate Arrays (FPGAs) are prefabricated integrated
circuits intended to be configured by the end user.  
FPGAs have grown significantly in terms of device features and gate counts.  
Figure~\ref{fig:fpga} illustrates an 
island-styled heterogeneous 
FPGA which consists of a two-dimensional 
array of Configurable Logic Blocks (CLBs) surrounded by peripheral I/O
blocks.  
The CLBs internally contain Look-Up Tables (LUTs) and Flip-Flips (FFs)
to implement logic.
Heterogeneous blocks such as Random Access Memories (RAMs) and 
digital signal processors (DSPs) are present. Other
hardened IP blocks (e.g., processors, SERDES, etc.) can also be included 
in the device.

Not obvious in Figure~\ref{fig:fpga} are the FPGA routing resources
which consist of wire segments and programmable switches.  
The programmable switches are used to connect wire segments to each other
and to other resources such as input and output pins on different blocks.
This information (i.e., the wire segments, programmable switches,
pins on blocks, etc.)
must be maintained during 
detailed routing
via
the \textit{Routing Resource Graph} (RRG) data structure.
The RRG can consume a significant amount
of memory which 
can negatively impact the router and the overall CAD flow.  In 
situations where insufficient memory is available, swap memory
will need to be used.  A large memory footprint will also have
negative consequences for cache.  

This observation is made in~\cite{2007a:Chin.S,2009a:Chin.S} and
the RRG memory footprint is reduced by relying on the FPGA
\textit{tiling}.
Tiling
means that ``copies'' of the same routing resources
are repeated over and over throughout the device.  
Different tiles may exist; e.g., the routing resources surrounding
CLBs might be different that those surrounding RAMs, DSPs or I/O blocks.
Nevertheless, when viewed as a graph, the RRG consists of 
many identical sub-graphs and that only a single copy of each sub-graph
(i.e., tile) is required in memory.  Tiles and resources can be expanded
as needed during FPGA routing.
This observation resulted in a $5$X to $13$X reduction in the memory
requirements for the RRG in~\cite{2007a:Chin.S,2009a:Chin.S}.  However, this approach
comes with a penalty.  Specifically, in~\cite{2007a:Chin.S,2009a:Chin.S} the 
required modifications to the routing algorithm resulted in 
$2.26$X runtime penalty for the time taken to perform detailed routing.

We propose an alternative view of the problem to compress the RRG.
Our approach benefits from FPGA tiling, but neither
requires nor relies upon it\footnote{We do not explicitly
identify tiles.
We also demonstrate compression when we ignore tiling.}.
Specifically, we focus on the \textit{adjacency information} in the 
RRG and compress only this portion of the RRG in two ways.
First, we apply delta
encoding and v-byte compression~\cite{2018a:Lemire.D} to reduce the size of the data
needed to store adjacency lists.  Second, we apply a sliding
window compression to avoid storing duplicate adjacency lists\footnote{Delta 
encoding and v-byte compression does
\textit{not} require tiling whereas the sliding window 
compression \textit{does benefit} from the repetition of resources
introduced by tiling.}.
Depending on the device, we can achieve as much as a $27$X reduction in 
the memory needed to store the adjacency information which translates
into as much as a $3.6$X reduction in the size of the RRG.
While 
this does not appear as significant as that reported in~\cite{2007a:Chin.S,2009a:Chin.S},
we find that our changes only slows down the router by an average of $\sim 20\%$ 
compared to the \textit{more that $2$X slowdown} reported 
in~\cite{2007a:Chin.S,2009a:Chin.S}.  Our proposed scheme
requires \textit{very few} lines of code to implement and
requires \textit{only 
insignificant and unintrusive changes} to the FPGA router.

Section~\ref{sec:background} 
provides background on RRGs.  We describe our algorithm in 
Section~\ref{sec:algo}.  Section~\ref{sec:router_mods} describes
changes to an FPGA router.  Different numerical results
are presented in Section~\ref{sec:results}.  Section~\ref{sec:conclusions}
presents our conclusions.

\section{Routing resource graphs}
\label{sec:background}

\begin{figure}[tb]
\begin{center}
\includegraphics[width=2.75in]{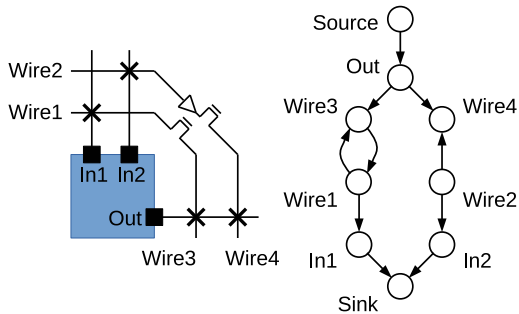} 
\end{center}
\vspace{-5mm}
\caption{Sample construction of the RRG from~\cite{2007a:Chin.S,2009a:Chin.S}.
Resources such as wires and pins are represented as nodes.  Edges exist
when it is possible to connect two resources together via a programmable
switch.}
\label{fig:fpga_rrg}
\end{figure}

The RRG is one piece of data which is maintained 
during routing and represents all of the physical resources inside
of the FPGA required to facilitate routing of signal nets.  This
information includes wires and pins as well as additional source 
and sink pins to model logically equivalences.  
A simple example of an RRG construction taken from ~\cite{2007a:Chin.S,2009a:Chin.S}
is illustrated in Figure~\ref{fig:fpga_rrg}.
Resources are represented as nodes in the RRG.  Programmable
connections (switches) between different resources are modeled as 
directed edges.  There can be a variety of different switches
available in the architecture; e.g., pass transistors or 
buffered switches with different resistances and capacitances.
Potential connections and switches form edges in the RRG and 
potentially consume significant memory.  An example of how
this information might be stored for a particular resource is
illustrated in Figure~\ref{fig:vpr_rr_node} in which adjacent
resources and the switch types used to make a connection are
stored as two vectors.
\begin{figure}
\begin{center}
\begin{algorithmic}
\State \textit{class RRGNode}
\State \textit{\{}
    \State \hspace{\algorithmicindent} \textit{// Adjacency information.}
    \State \hspace{\algorithmicindent} $std::vector<short> switches;$
    \State \hspace{\algorithmicindent} $std::vector<int> edges;$
    \State \hspace{\algorithmicindent} \textit{// Additional resource details.}
    \State \hspace{\algorithmicindent} \textit{...}
\State \textit{\};}
\end{algorithmic}
\caption{Sample of adjacency information stored in an RRG node.  The
structure is shown similar to the structure found in the
VPR~\cite{1999a:Betz.V} tool.}
\label{fig:vpr_rr_node}
\end{center}
\end{figure}

Additional information is stored in each node, but we don't 
consider it.  This includes the physical location,
span, occupancy, capacity, resistance and capacitance of the
resource.  However, we are only interested in the adjacencies;
it appears there is significant potential to reduce the 
memory requirements by focusing only on this information.
As illustrated in Figure~\ref{fig:vpr_rr_node},
each edge in the RRG would require one \texttt{int} and one
\texttt{short}, consuming a minimum of \texttt{6 bytes} per edge.
Our aim is to either \textit{compress} or \textit{eliminate}
as much of this information as possible without losing any RRG
details.

\section{Our algorithm}
\label{sec:algo}

\subsection{Delta encoding and v-byte compression}
\label{sec:delta}

Nodes in the RRG are typically created in a regular localized 
manner; e.g., 
given an $(x,y)$ coordinate in the FPGA, all routing resources
around that $(x,y)$ coordinate are created at the same time.
Further, these routing resources are typically connected 
locally.  This implies that two RRG nodes $i$ and $j$ near the
same $(x,y)$ coordinate will have similar integer identifiers
and this fact can be exploited.

Figure~\ref{fig:adjacent_example} shows an adjacency list for an RRG
node with 7 adjacencies which required \texttt{28} bytes minimum
(we don't consider the switches in our explanation).
\begin{figure}
\begin{center}
\begin{tabular}{c}
\begin{tabular}{rl}
Adjacency List:     & 44,62,387,401,414,430,910 \\
Delta List:   & 44,18,325,14,13,16,480 \\
V-Bytes List: &
        $\underline{\textrm{AC}}$,
        $\underline{\textrm{92}}$,
        $\underline{\textrm{02}} \, \underline{\textrm{C5}}$,
        $\underline{\textrm{8E}}$,
        $\underline{\textrm{8D}}$,
        $\underline{\textrm{90}}$,
        $\underline{\textrm{03}} \, \underline{\textrm{E0}}$ \\
\end{tabular}
\end{tabular}
\end{center}
\caption{RRG node adjacency list: (a) Ids of adjacent resources
via their integer ids; (b) Same data using delta encoding;
(c) The data (in hex) using v-byte
to compress the deltas into as few bytes as possible.}
\label{fig:adjacent_example}
\end{figure}
Figure~\ref{fig:adjacent_example} also shows the same information,
but with delta encoding.  With delta encoding, the deltas 
between consecutive sorted integers is always a \textit{smaller} positive
integer. 
It is not necessary to use \texttt{4 bytes} to store each 
adjacency and this is even more true with delta encoding.
We can apply v-byte
encoding to \textit{compress} the adjacencies into even fewer
bytes and only use as many bytes as required.  The compressed
v-byte encoded adjacency list is also shown in Figure~\ref{fig:adjacent_example}.
The integer ids of adjacent resources can be reduced from
7 integers (\texttt{28 bytes}) to only \texttt{9 bytes}.  It
is also possible to compress the switch information 
into bytes effectively.

The complexity of the compression is $O(n \log n)$ where $n$ is the 
length of the adjacency list due to the need to sort the adjacency
lists.  Deltas and compression is achieved with a single pass over
the adjacency list.
Compression pseudocode is given in Figure~\ref{fig:adjacent_compress}.
Adjacency lists are compressed as they are created to avoid using
large amounts of memory to represent the RRG.
Numerical results demonstrate the effectiveness of delta encoding and
v-bytes compression.
\begin{figure}
\begin{tabular}{c}
\begin{minipage}{4in}
\small
\begin{algorithmic}
\State \textbf{Input:} $edges$ \textit{// vector of integer adjacencies}
\State \textbf{Output:} $compressed$ \textit{// vector of compressed byte adjacencies}
\State $std::sort( edges.begin(), edges.end() );$
\State $last \gets 0;$
\State $compressed.erase( compressed.begin(), compressed.end() );$
\For{($i = 0; i < edges.size(); i \textrm{++} )$}
    \State $diff = edges[i]-last;$

    \State $s.erase( s.begin(), s.end() );$
    \Loop
        \State $s.push\_front( diff\%128 );$
        \If{$diff < 128$} 
            \State $break;$ 
        \Else 
            \State $diff = diff / 128;$ 
        \EndIf
    \EndLoop
    \State $s.back() += 128;$
    \State $compressed.push\_back( s.begin(), s.end() );$
    \State $last \gets edges[i];$
\EndFor
\end{algorithmic}
\end{minipage}
\end{tabular}
\caption{Pseudocode for adjacency list compression during RRG creation.}
\label{fig:adjacent_compress}
\end{figure}

\subsection{Sliding window compression}
\label{sec:tile}

Although we don't specifically require tiling, it is very likely that
there are repeated patterns in the routing graph due to tiling.
Figure~\ref{fig:tiling} shows the adjacencies for two resources
in different physical regions of the FPGA after delta encoding.
\begin{figure}
\begin{center}
\begin{tabular}{rcl}
Node \#373 & : &
   334,10,15,4,39,451,23,6 \\
Node \#8564 & : &
    8525,10,15,4,39,451,23,6 \\
\multicolumn{3}{c}{(a)} \\
Node \#373 & : &
    334,10,15,4,39,451,23,6 \\
Node \#8564 (referenced) & : & 8525,-373\\
\multicolumn{3}{c}{(b)} \\
\end{tabular}
\end{center}
\caption{Two different RRG nodes: (a) Adjacency lists for each
node after delta encoding;
(b) Stored information in which the second node has a 
reference to the other node for delta values.}
\label{fig:tiling}
\end{figure}
It is clear that the repeated patterns of connections are made
clear by the use of delta encoding.  Consequently, the storage
of \textit{both} adacency lists is not required.
During RRG creation, adjacency patterns (the deltas) are hashed
and if RRG nodes with identical deltas are encountered, their
adjacency lists are not stored explicitly (only a reference is
recorded to the other RRG node).

\subsection{Node renumbering}

\begin{figure*}
\begin{center}
\begin{tabular}{ccc}
\includegraphics[width=2.25in]{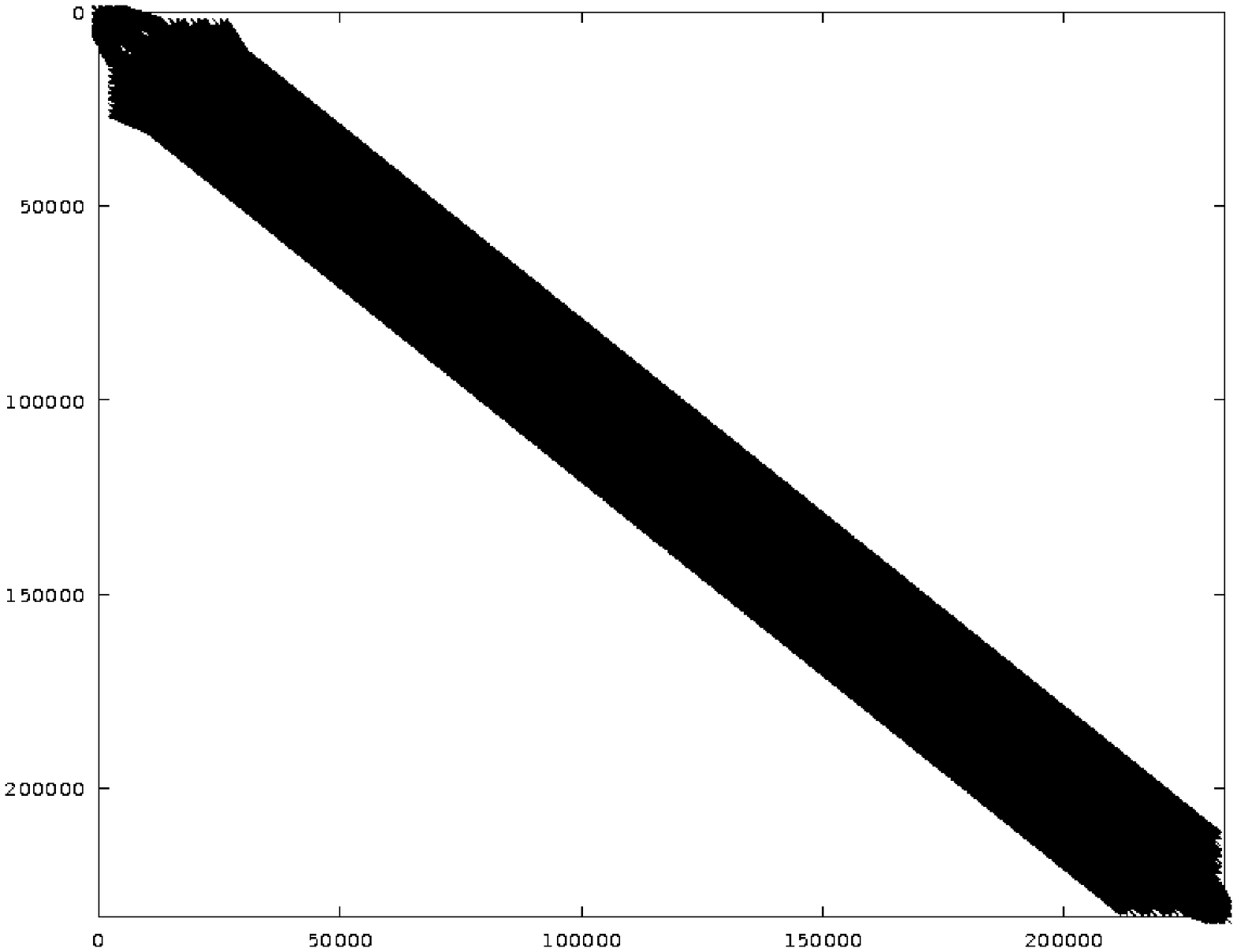}  &
\includegraphics[width=2.25in]{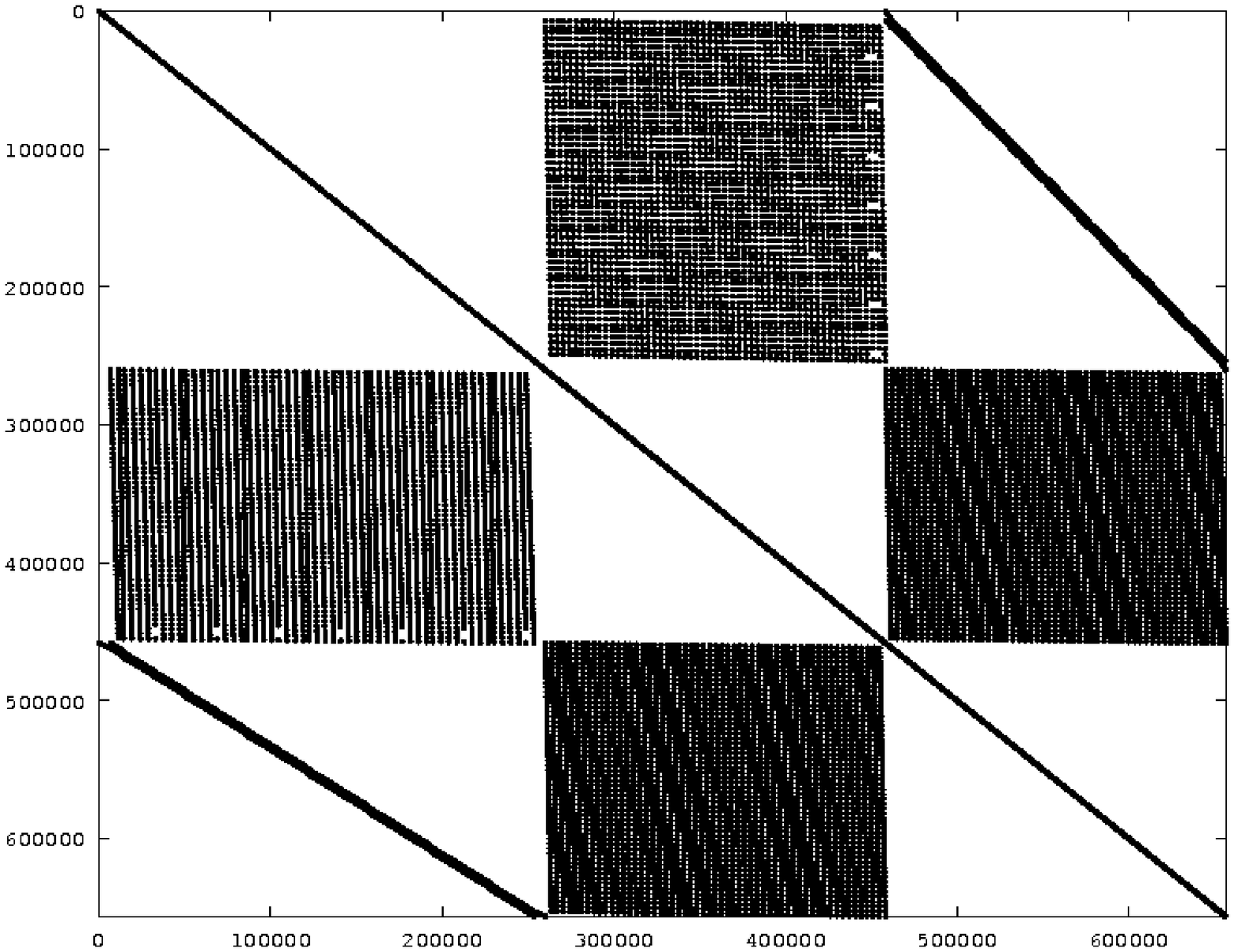}  &
\includegraphics[width=2.25in]{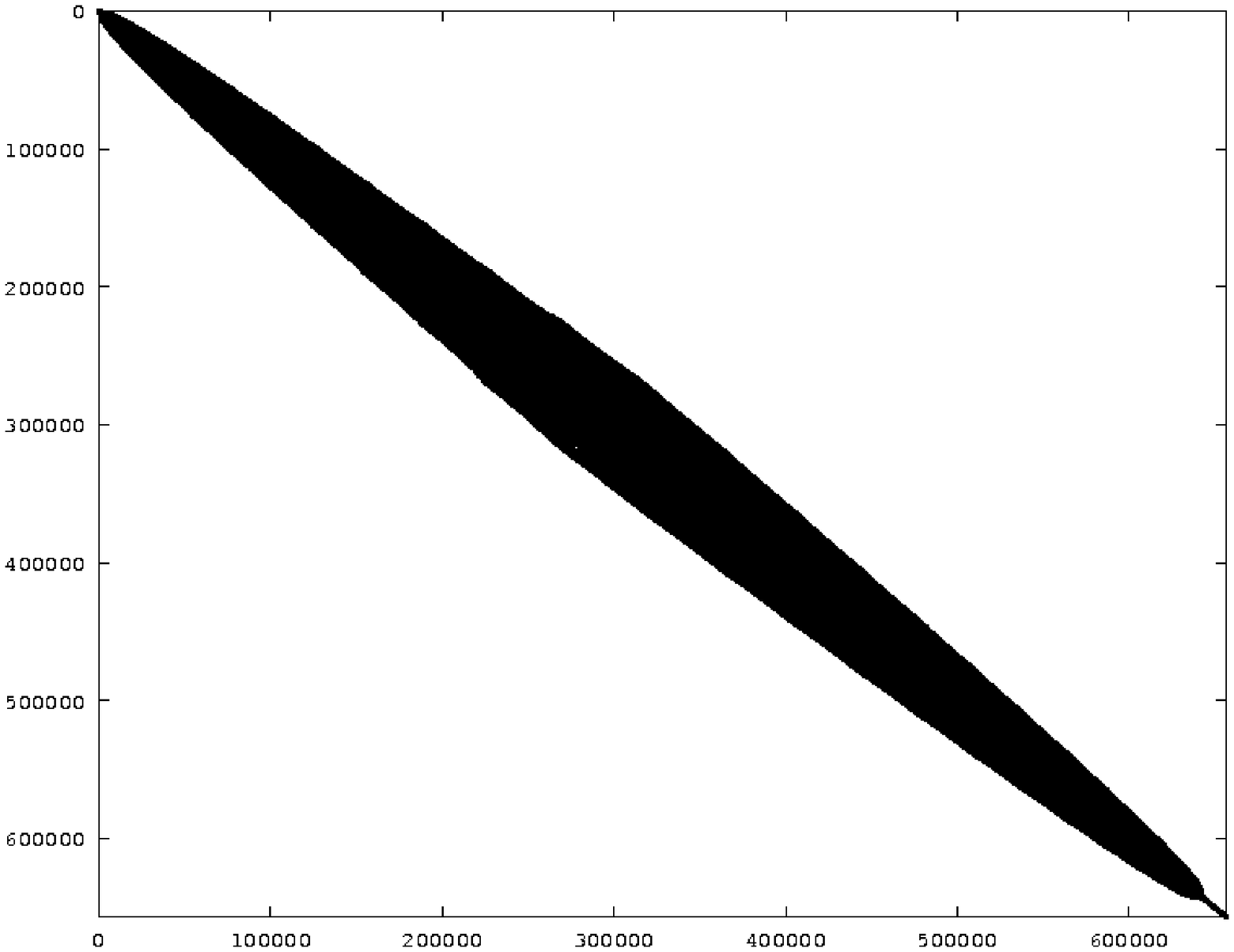}  \\
(a) & (b) & (c) \\
\end{tabular}
\caption{Sparsity patterns for the RRG adjacencies:
(a) A homogeneous FPGA device; (b) A heterogeneous FGPA device;
(c) The heterogeneous FPGA device (b) with RRG nodes renumbered.}
\label{fig:adjacency_matrices}
\end{center}
\end{figure*}

We also consider renumbering RRG nodes.  Since adjacencies can
be viewed as a matrix, we consider the use of matrix reordering
techniques.  It is typical that RRGs are created such that
physically close resources are created with similar identifiers
and connected.  For example, 
Figure~\ref{fig:adjacency_matrices}(a) shows the RRG for a 
homogeneous architecture when viewed as an adjacency matrix.
In Figure~\ref{fig:adjacency_matrices}(a), the clustering of adjacencies 
around the diagonal indicates connected nodes have similar identifiers
which benefits the use of delta encoded lists.

The situation is slighly different for a heterogeneous device whose
RRG is depicited in Figure~\ref{fig:adjacency_matrices}(b).  Entries
are not always clustered around the diagonal, although blocks appear
off diagonal.  The use of delta encoding could be impaired.
Figure~\ref{fig:adjacency_matrices}(c) shows the same
RRG as Figure~\ref{fig:adjacency_matrices}(b),
but with RRG nodes renumbered through \texttt{RCM} matrix 
reordering~\cite{1981a:George.A}.  The bandwidth reduction obtained
through matrix reordering can potentially result in smaller deltas
which will benefit compression.
However, we note that renumbering RRG nodes can potentially ``hide''
any tiling within the device and negatively impact the previously 
described windowing strategy.

\section{Router modifications}
\label{sec:router_mods}

Typical FPGA routers are based on path-finding 
algorithms~\cite{1995a:McMurchie.L,1999a:Betz.V} which use graph
search algorithms (e.g., BFS or $A^{*}$).
Each sign net is first routed ignoring other nets and 
then a loop is entered as long as there are overused RRG resources
due to multiple nets requesting the same RRG resources.
In each iteration, each net is ripped up and re-routed and
effort is made to avoid reusing overused
resources.  
Each net is then routed using some sort of graph search algorithm
and a salient feature of these algorithms is the need for
``neighborhood expansion''
which
required looping over the adjacencies of an RRG node.  

In our case, these adjacencies are either encoded and compressed
or found via a reference to another node (in which
the information is also encoded or compressed).  Prior
to expanding neighbors, we must \textit{derefernce, decompress and
decode} the adjacency lists.
This is the \textit{only modification} needed to the router.
The decompression and decoding is linear and is done as
shown in Figure~\ref{fig:routing_algorithm}.  
\begin{figure}
\begin{tabular}{c}
\begin{minipage}{4in}
\small
\begin{algorithmic}
\State \textbf{Input:} $compressed$ 
    \textit{// vector of compressed byte adjacencies}
\State \textbf{Output:} $edges$ 
    \textit{// vector of integer adjacencies}
\State $edges.erase( edges.begin(), edges.end() );$
\State $last \gets 0;$
\For{($i = 0; i < compressed.size(); )$}
    \State $diff \gets 0;$
    \Loop
        \If{$compressed[i] < 128$}
            \State{$diff = 128*diff + compressed[i]$;}
            \State{$i\textrm{++};$}
        \Else
            \State{$diff = 128*diff + compressed[i]-128$;}
            \State{$i\textrm{++};$}
            \State{$break;$}
        \EndIf
    \EndLoop
    \State{$edges.push\_back( diff-last );$}
    \State{$last \gets edges.back();$}
\EndFor
\end{algorithmic}
\end{minipage}
\end{tabular}
\caption{Pseudocode for decompression neighborhood expansion.  
If an adjacency list is referenced to another node (not shown), 
then deltas should be extracted from the other node.}
\label{fig:routing_algorithm}
\end{figure}

\section{Numerical Results}
\label{sec:results}

We modified \texttt{VPR4.3}~\cite{1999a:Betz.V} and
\texttt{VTR}~\cite{2012a:Rose.J}\footnote{We consider
\texttt{VPR4.3} and homogeneous FPGAs for 
some comparison to~\cite{2007a:Chin.S,2009a:Chin.S}.
We used the lastest version of \texttt{VTR} from
\texttt{github} to consider heterogeneous FPGAs.}
to examine compressed
RRG storage requirements and router runtimes.

\subsection{Memory reductions}

\begin{figure}
\begin{center}
\begin{tabular}{c}
\includegraphics[width=3.50in]{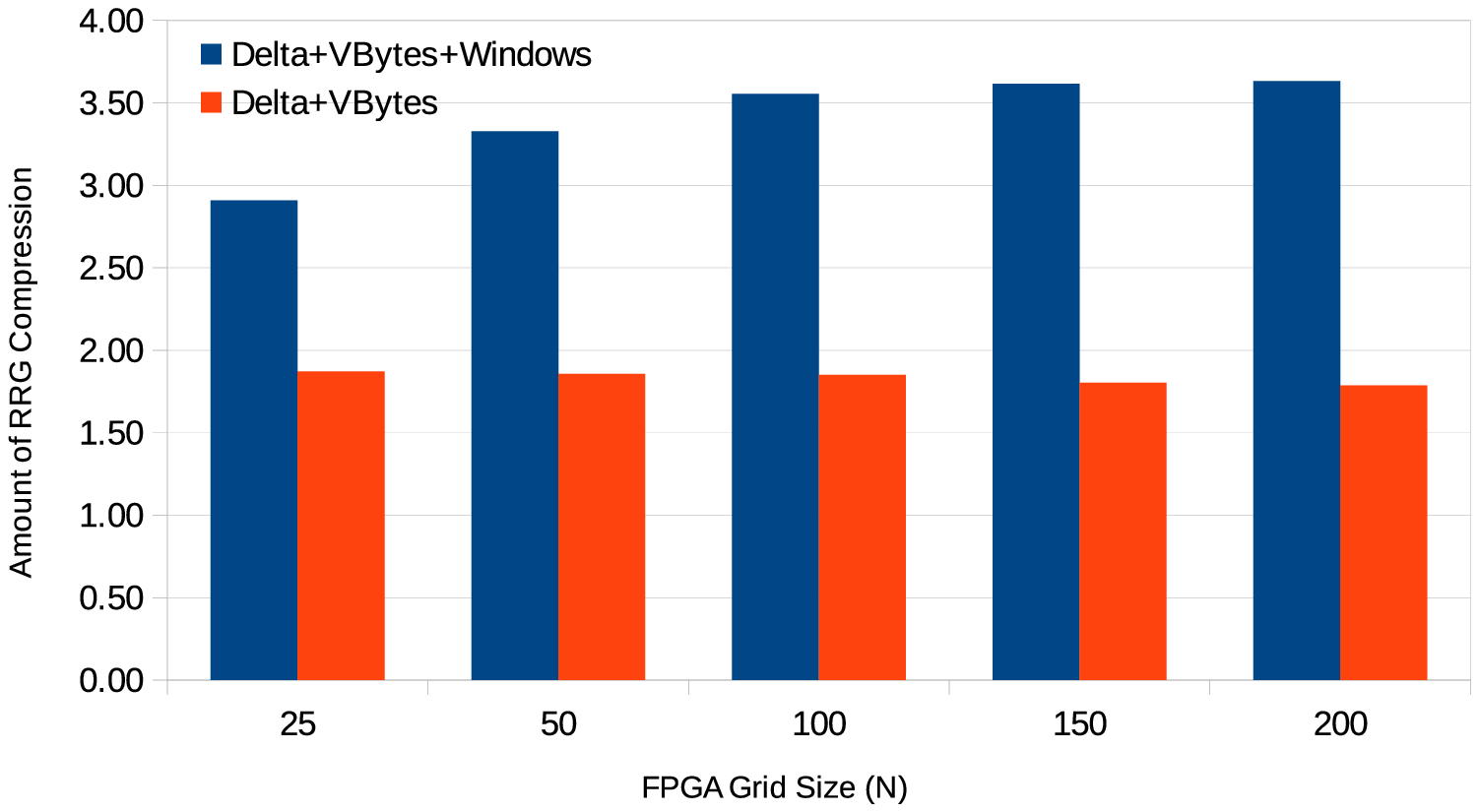} \vspace{-2mm} \\
(a) \\
\includegraphics[width=3.50in]{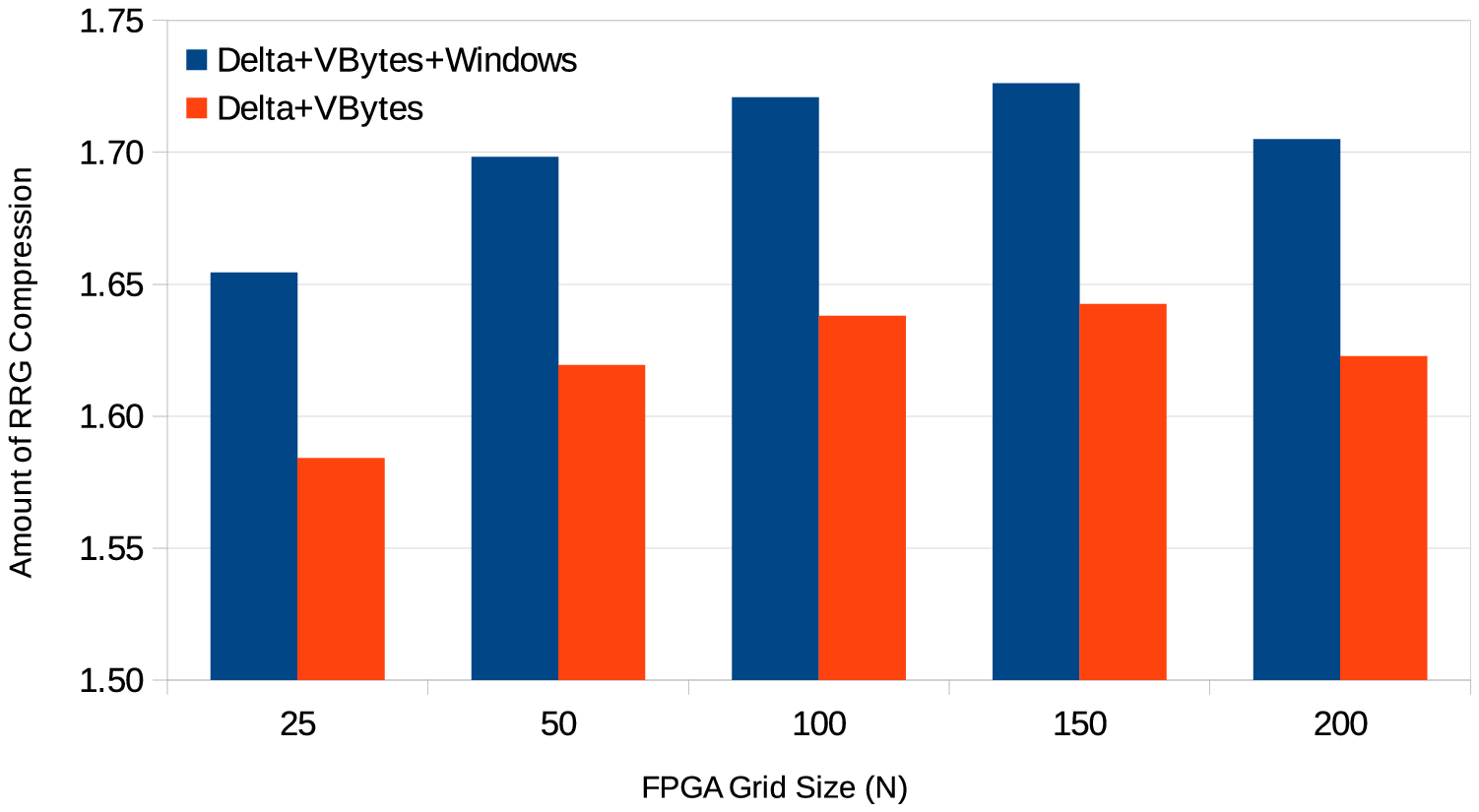} \vspace{-2mm} \\
(b) \\
\end{tabular}
\end{center}
\vspace{-2mm}
\caption{RRG compression results for different sized FPGAs: 
(a) homogeneous architectures using \texttt{VPR4.3}; 
(b) heterogeneous architectures using \texttt{VTR}.}
\label{fig:mem_vary_n}
\end{figure}

Figure~\ref{fig:mem_vary_n}(a) shows the RRG compression for 
different FPGA grid sizes using \texttt{VPR4.3} in which 
each CLB contained 10 LUT/FF pairs with 22 inputs and 10 outputs, 
respectively.  The channel width was fixed at 150 wires of length 4. 
We found that adjacency
information alone consumed $\sim 76\%$ of the RRG memory 
indicating this is a ``valid target'' for compression.
With all compression options, we see a reduction of $2.9$X $\sim$ $3.6$X 
in the RRG size (a $7.8$X $\sim$ $27.0$X reduction the storage
requirements if only the adjacency information is considered).
With only delta encoding with v-byte compression, 
Figure~\ref{fig:mem_vary_n}(a) shows compressions of $1.8$X $\sim$ $1.9$X.
While not as substantial, this compression is \textit{readily available}
without relying on the FPGA tiling.

Figure~\ref{fig:mem_vary_n}(b) shows the same results for a heterogeneous
architectures created using \texttt{VTR}\footnote{The  
heterogeneous FPGAs included DSP and RAM blocks and were created 
using the \texttt{k6\_frac\_N10\_mem32K\_40nm.xml} architecture file.}.
Here, the adjacency information consumed $\sim 68\%$ of the total RRG
memory.
Figure~\ref{fig:mem_vary_n}(b) shows a savings of $1.58$X $\sim$ $1.62$X
using delta encoding and v-bytes compression.  With all options, we 
see only a modest improvement of $1.65$X $\sim$ $1.73$X.  Interestingly, 
with these heterogenous architectures, windowing does not seem to 
significantly improve the RRG compression.

We do not include compression results obtained using RRG node renumbering.
Our investigation thus far into this technique has not been useful for achieving improved
compression results.  Specifically, we found that node renumbering was 
effective at reducing the magnitudes of the delta values, but only lead
to minor improvement in the compression.  However, finding identical 
patterns of delta values through windowing became less effective.  The
net result was little to no change in the overall compression ratios.

\subsection{Router impact}
Routing results are deterministic so we only consider the impact on
the routing runtime due to the need to constantly decompress RRG
adjacencies.
Figure~\ref{fig:routing_results}(a) shows the results on a set of 
designs run through \texttt{VPR4.3}.  Each CLB consisted
of a single LUT/FF pair and all wire segments are length 1.   
Designs were mapped to the smallest FPGA into which they would fit.
Figure~\ref{fig:routing_results}(a) shows the router runtime is
impacted by only $14\%$ on average with a maximum penalty of $20\%$
which compares very favorably to the runtime impact 
mentioned in~\cite{2007a:Chin.S,2009a:Chin.S}.  Figure~\ref{fig:routing_results}(b)
shows the same results for a set of heterogeneous designs run through
\texttt{VTR} into a heterogenous FPGA.  Here we see a runtime impact
of $20\%$ on average with a maximum penalty of $25\%$.
\begin{figure}[th]
\begin{center}
\begin{tabular}{c}
\begin{tabular}{|c|c|} \hline\hline 
 \textbf{Runtime Ratio}  & \textbf{RRG Memory Reduction} \\ 
 \textbf{Min/Max/Avg}    & \textbf{Min/Max/Avg} \\ \hline\hline
 1.08/1.20/1.14  & 1.56/1.72/1.62 \\ \hline\hline
\end{tabular} \\
(a) \\
\begin{tabular}{|c|c|} \hline\hline 
 \textbf{Runtime Ratio}  & \textbf{RRG Memory Reduction} \\ 
 \textbf{Min/Max/Avg}    & \textbf{Min/Max/Avg} \\ \hline\hline
 1.12/1.25/1.20  & 1.30/1.74/1.48 \\ \hline\hline
\end{tabular} \\
(b) \\
\end{tabular}
\end{center}
\normalsize
\caption{Impact of RRG compression on detailed router runtimes; (a)
homogeneous designs mapped to \texttt{VPR4.3}; (b) heterogeneous
designs mapped to \texttt{VTR}.}
\label{fig:routing_results}
\end{figure}
Figure~\ref{fig:routing_results} also shows the RRG compressions 
achieved for these additional FPGA devices to demonstrate memory
savings.

\section{Conclusions}
\label{sec:conclusions}

We have presented several simple ideas to compress the RRGs 
used by FPGA detailed routers.   Our ideas are extremely 
easy to implement and focus on viewing the RRG as an adjacency
graph.  The compression achieved is reasonable and appears to
\textit{not significantly impact} detailed router runtimes. 

Our compression and decompression implementations were straightforward
and the impact on router runtime could possibly be reduced by using
more efficient decompression~\cite{2018a:Lemire.D}.  
More investigation of compression on heterogeneous architectures
seems worthwhile.

\bibliographystyle{abbrv}

\begin{thebibliography}{1}

\bibitem{1999a:Betz.V}
V.~Betz, J.~Rose, and A.~Marquardt.
\newblock {\em Architecture and CAD For Deep-Submicron FPGAs}.
\newblock Kluwer Academic Publishers, 1999.

\bibitem{2009a:Chin.S}
S.~Y.~L. Chin and S.~J.~E. Wilton.
\newblock Static and dynamic memory footprint reduction for {FPGA} routing
  algorithms.
\newblock {\em TREATS}, 1(4):1--20.

\bibitem{2007a:Chin.S}
S.~Y.~L. Chin and S.~J.~E. Wilton.
\newblock Memory footprint reduction for {FPGA} routing algorithms.
\newblock In {\em Proc.\ FPT}, pages 1--8, 2007.

\bibitem{1981a:George.A}
A.~George and J.~Liu.
\newblock {\em Computer Solution of Large Sparse Positive Definite Matrices}.
\newblock Prentice Hall, 1981.

\bibitem{2018a:Lemire.D}
D.~Lemire, N.~Kurz, and C.~Rupp.
\newblock Stream {VB}yte: Faster byte-oriented integer compression.
\newblock {\em Information Processing Letters}, 130:1--6, February 2018.

\bibitem{1995a:McMurchie.L}
L.~McMurchie and C.~Ebeling.
\newblock Pathfinder: {A} negotiation-based performance-driven router for
  {FPGA}s.
\newblock In {\em Proc.\ FPGA}, pages 111--117, 1995.

\bibitem{2012a:Rose.J}
J.~Rose, J.~Luu, C.~W. Yu, O.~Densmore, J.~Goeders, A.~Somerville, K.~B. Kent,
  P.~Jamieson, , and J.~Anderson.
\newblock The {VTR} project: architecture and {CAD} for {FPGA} from verilog to
  routing.
\newblock In {\em FPGA}, pages 77--86, 2012.

\end{thebibliography}

%
%
%
%
%
%
%
%

\end{document}